\documentclass[12pt]{iopart}
\usepackage[dvips]{graphics}
\usepackage{color}

\newcommand{\beq}{\begin{equation}}
\newcommand{\eeq}{\end{equation}}
\newcommand{\bma}{\begin{math}}
\newcommand{\ema}{\end{math}}
\newcommand{\beqa}{\begin{eqnarray}}
\newcommand{\eeqa}{\end{eqnarray}}

\def\expect#1{\langle\, #1\, \rangle}

\def\opone{\le\textbf{}\textbf{}avevmode\hbox{\small1\kern-3.8pt\normalsize1}}

\newcommand{\be}[1]{     \begin{eqnarray} \mbox{$\label{#1}$}   }

\newcommand{\ee}{\end{eqnarray}}
\newcommand{\pref}[1]{(\ref{#1})}

\newcommand\half{\frac 1 2 }

\newcommand\ket [1] {|#1 \rangle }

\begin{document}

\title{Quantum Hall Circle}

\author{E.J. Bergholtz$^1$ and A. Karlhede$^2$}

\address{
$^1$Max Planck Institute  for the Physics of Complex Systems,\\ Noethnitzer Str. 38, 01187 Dresden, Germany\\
$^2$Department of Physics, Stockholm University, \\
AlbaNova University Center, SE-106 91 Stockholm, Sweden}
\eads{\mailto{ejb@pks.mpg.de} and
\mailto{ak@physto.se}}

\date{\today}

\begin{abstract}
We consider spin-polarized electrons in a single Landau level on a cylinder as the circumference of the cylinder goes to infinity. 
This gives a model of interacting electrons  on a circle where the momenta of the particles are restricted and there is no kinetic energy. 
Quantum Hall states are exact ground states for appropriate short range interactions, and there is a gap to excitations. 
These states develop adiabatically from this one-dimensional quantum Hall circle to the bulk quantum Hall states and further on into the 
Tao-Thouless states as the circumference goes to zero. For low filling fractions a gapless state is formed which we suggest is connected 
to the Wigner crystal expected in the bulk. 
\end{abstract}


\maketitle

\section{Introduction}\label{introduction}

The quantum Hall (QH) problem of interacting spin-polarized electrons in a single Landau level has been studied on a cylinder as the circumference $L$  varies \cite{Haldane94,bk1,Lee05,bk2} and the hamiltonian has been diagonalized in the limit $L\rightarrow 0$ \cite{bk2}. The ground state is, at filling factor $\nu=p/q$, $q$-fold degenerate and there is a gap to excitations. Domain walls between the degenerate ground states are quasiparticles with charge $\pm e/q$. For the Laughlin fractions, the ground states are those introduced by Tao and Thouless in 1983 \cite{TaoThouless83}. For odd $q$, these ground states are adiabatically connected to bulk abelian QH states and are the limits of the Laughlin \cite{Laughlin83} and Jain \cite{jain} wave functions for filling factors where these exist. Details about, and extensions of, this so-called Tao-Thouless limit can be found in Ref. \cite{bk3,bk4}. 

In this article we study the limit opposite to the Tao-Thouless limit, namely  when the circumference of the cylinder goes to infinity keeping the filling fraction and the number of particles, $N$, and hence the area of the system, fixed. When $L\rightarrow \infty$, the distance along the cylinder between the single-particle states goes to zero and the cylinder becomes a narrow hoop. We show that the QH problem becomes a model of electrons on a circle, where the range of the momenta, $k$, of the electrons are restricted by the filling factor and the number of particles, $0\le k\le N/\nu -const$.  The hamiltonian consists only of an  effective electron-electron repulsion; there is no kinetic term. It is the restriction on the particles momenta that makes the model non-trivial and connects it to the bulk QH problem. We call this one-dimensional limit the quantum Hall circle. The Laughlin state at $\nu=1/q$ is the unique ground state and there is a gap to excitations on the cylinder for any $L$, including the QH circle, for the short range interaction \cite{haldanebook,trugman, pokrovsky}. 
Thus,  the ground state for the quantum Hall circle is adiabatically connected to the bulk QH state as $L$ decreases, and eventually to the Tao-Thouless state as $L\rightarrow 0$.
We believe this adiabatic continuity can be generalized to other QH states which  are ground states of a known interaction, including non-abelian states such as the Moore-Read \cite{mr} and the Read-Rezayi states \cite{readrezayi}. For low filling factors, we argue that there is a gapless  ground state reminiscent of the Wigner crystal expected in the bulk QH system \cite{w0}.

The  QH circle, obtained here as a limit of the QH problem, is identical to the toy model, introduced by Dyakonov  \cite{Dyakonov}.  He identified a crystal state for small filling factors and suggested a Laughlin state for larger $\nu$. 
Some aspects of the circle limit of the QH problem have been considered earlier by  Rezayi and Haldane \cite{Haldane94}, and by Langmann and Nordblad \cite{Langmann}. 

The article is organized as follows. In Sec. \ref{circle} the circle limit is derived. Quantum Hall states, in particular the Laughlin fractions with a short range interaction, are discussed in Sec. \ref{qh}, with details given in an Appendix, and in Sec. \ref{crystal} a localized Wannier-like basis is used to discuss the gapless crystal state. Finally, in Sec. \ref{disc} we discuss our findings.

\section{Circle }\label{circle}

Here we derive the one-dimensional model, the quantum Hall circle, starting from interacting spin-polarized electrons in a single Landau level. The hamiltonian is
\be{ham1}
H=\frac 1 2 \int d{\bf r} d{\bf r}^\prime :( \rho ({\bf r}) - \bar \rho )V(|{\bf r}-{\bf r}^\prime|) (\rho ({\bf r}) - \bar \rho ): ,
\ee
where $V(|{\bf r}|)$ is some electron-electron interaction, $ \rho ({\bf r})=\hat \Psi^\dagger ({\bf r})\hat \Psi ({\bf r})$ is the electron 
density operator ($\hat \Psi ({\bf r})$ being the electron annihilation operator),  $\bar \rho $ is a compensating background charge 
and $: \ :$ denotes normal ordering. 

We consider a cylinder with circumference $L$ and coordinates ${\bf r}=(x,y)$, $x$ going around the cylinder.  
A basis of  single-particle states, obeying periodic boundary conditions $\psi(x+L)=\psi(x)$, in Landau level $p$ are
\begin{eqnarray}\label{psicyl}
\psi_{pk}(\mathbf{r})&=&(\sqrt \pi 2^p p! L)^{-1/2} H_p(y+2\pi k/L)e^{2\pi ikx/L}e^{-(y+2\pi k/L)^{2}/2}\  .
\end{eqnarray}
Here, $H_p$ is the $p$th Hermite polynomial and $k$ is an integer corresponding to the conserved momentum $2\pi k/L$ in the $x$-direction; 
$k$ also gives the position along the cylinder: $\psi_{pk}$ is a gaussian centered at $y=-2\pi k/L$. Lengths are given in units of the magnetic length $\ell=\sqrt{\hbar c/eB}$.

Restricting to a single Landau level $p$, we have, dropping the Landau level index $\psi_k=\psi_{pk}$,
\begin{eqnarray}
\rho ({\bf r}) =\sum _{k,l} \psi_{k}^* ({\bf r}) \psi_{l} ({\bf r}) c^\dagger_k c_l \  ,
\end{eqnarray}
where $c^\dagger _k$ creates an electron with momentum $k$, $\{c^\dagger_k,c_l\}=\delta_{kl}$. In momentum space the hamiltonian becomes, 
dropping the background $\bar \rho$,
\be{hammom}
H=\sum_{|m| <k}V_{km}\sum_i c^\dagger_{m+i} c^\dagger_{k+i} c_{m+k+i} c_i \  \ ,
\ee
where
\be{matrixelements}
V_{km}= \int d{\bf r} d{\bf r}^\prime \psi_{m}^*({\bf r})\psi_{k}^*({\bf r}^\prime)V(|{\bf r}-{\bf r}^\prime|) \psi_{m+k}({\bf r}^\prime)\psi_{0}({\bf r})
- (k \leftrightarrow m) \ \ .
\ee
($V_{km}$ are real and $V_{k,-m}=V_{km}$, hence $H$ is hermitean. The interaction is assumed to be periodic $V(|{\bf r}+L\hat x|)=V(|{\bf r}|)$.)

So far this is just the standard problem of interacting electrons in a single  Landau level $p$. We now introduce the angular variable $\varphi=2\pi x/L$ 
and take the  $L \rightarrow \infty$ limit of the one-electron states
\begin{eqnarray}\label{psilimit}
\psi_{k} \rightarrow (\sqrt \pi 2^p p! L)^{-1/2} e^{ik \varphi}H_p(y)e^{-y^{2}/2} \ .
\end{eqnarray}
The distance  between the states along the cylinder has shrunk to zero, and the $y$-dependence is just an overall  $k$-independent factor which 
can hence be integrated out in \pref{ham1} leading to a theory of electrons on a circle with an effective interaction $\tilde V (\varphi)$, {\it cf} \pref{hamcont1d} below. 
For given filling fraction $\nu$, the momenta $k$ are restricted to a finite range; this motivates dropping the $k$-dependence in the gaussian factor in \pref{psilimit}. 
The hamiltonian is of course 
still given by \pref{hammom} but now with matrix elements
\begin{eqnarray}\label{v1d}
V_{km}&=&\frac 1 {(2\pi)^2} \int _0^{2\pi} d\varphi d\varphi^\prime (e^{ i m(\varphi^\prime-\varphi)}-e^{ i k(\varphi^\prime-\varphi)})\tilde V(\varphi-\varphi^\prime) \nonumber  \nonumber \\
&=& \frac 1 {2\pi} \int _0^{2\pi} d\varphi  (e^{ i m\varphi}-e^{ i k\varphi})\tilde V(\varphi) \ ,
\end{eqnarray}
where
\begin{eqnarray}\label{veff}
\tilde V(\varphi)&=&   \frac 1 {(2^{p}p!)^2\pi}  \int _{-\infty}^{\infty} dy dy^\prime V(\sqrt{(\frac {L\varphi} {2\pi})^2+(y-y^\prime)^2}) \nonumber \\
&\times& (H_p(y)H_p(y^\prime))^2 e^{-y^2-y^{\prime 2}}
\end{eqnarray}
is the effective interaction on the circle.  
For the periodic Coulomb interaction in the lowest Landau level, $V({\bf r})=\sum_n(1/|{\bf r}+n L\hat x |)$ and $p=0$, we find
\be{}
\tilde V(\varphi) &=&\frac {2^{3/2}} {\sqrt \pi} \sum_n e^{(L(\varphi+2n\pi)/4\pi)^2}K_0((\frac L {4\pi} (\varphi+2n\pi))^2) \nonumber \\
&=&\frac { 8 \pi} {L} \sum_n \frac 1 {|\varphi+2n\pi|}+ {\cal O} ((L\varphi)^{-3}) \ ,
\ee
where $K_0$ is the modified Bessel function of the second kind. 

Introducing the one-dimensional electron operators
\be{}
\hat \psi (\varphi) =\frac 1 {\sqrt {2 \pi}} \sum _{k=-\infty}^{\infty} e^{ i k \varphi} c_k \ \ ,
\ee
which obey standard fermionic commutation relations
\be{}
\{\hat \psi(\varphi),\hat \psi^\dagger(\varphi^\prime)\}=\delta(\varphi-\varphi^\prime) \,   , \ \ \  \delta(\varphi+2\pi) =\delta (\varphi)  \,  ,
\ee
the hamiltonian \pref{hammom}, or equivalently \pref{ham1} without $\bar \rho$, becomes
\be{hamcont1d}
H&=&\frac 1 2   \int_0^{2\pi}d\varphi d\varphi^\prime : \rho (\varphi) \tilde V(\varphi-\varphi^\prime) \rho (\varphi ^\prime):  \\
&=& \half \int_0^{2\pi}d\varphi  \rho (\varphi) \int_0^{2\pi} d\varphi^\prime  \{ \tilde V(\varphi-\varphi^\prime) \rho (\varphi ^\prime)-  \tilde V(\varphi)\}  \nonumber  \ ,
\ee
where 
\be{}
\rho (\varphi) =  \hat \psi^\dagger(\varphi )\hat \psi(\varphi) \ . 
\ee
This hamiltonian could also have been derived directly from \pref{ham1} using \pref{psilimit}. 
In obtaining the one-dimensional theory \pref{hamcont1d} it is crucial that the effective interaction $\tilde V(\varphi)$ in (\ref{v1d},\ref{veff}) is independent of the momentum index $k$. 
This is the case provided the limit is taken as in \pref{psilimit}, {\it ie} provided the $k$-dependence in $e^{-(y+2\pi k/L)^2/2}$ is dropped. This is consistent, in the sense that 
\pref{hamcont1d} is indeed the $L\rightarrow \infty$ limit of the hamiltonian \pref{ham1}, as long as the interaction $V(r)$ is regular enough. This is  the case for, {\it eg}, the Coulomb interaction $V(r)=1/r$ but not for a short-range interaction $\nabla ^{2s} \delta ^2 ({\bf r})$ as will be discussed below.

The hamiltonian in \pref{hamcont1d} describes simply a system of interacting electrons on a circle  without a kinetic term.
As such it can of course be diagonalized, as has been noted before \cite{Langmann}. The energy eigenstates are simply the position eigenstates 
\be{psie}\ket{\Psi_E}=\prod_i \hat \psi^\dagger(\varphi_i) \ket{0}\ee with energies
\be{energy}
E=\half \{\sum_{j,k} \tilde V(\varphi_k-\varphi_j)- \sum_j \tilde V(\varphi_j)\}  .\ee
However,  this exact solution is presumably not relevant for the quantum Hall problem since 
we have not implemented a specific filling factor. Referring back to the full quantum Hall problem, before taking the limit, 
we see that the positions of the single-electron states along the cylinder are given by their momenta, hence the filling factor $\nu$ is implemented by 
imposing a restriction in momentum space. For $N$ electrons at  $\nu=1/q$, the allowed momentum states are $0\le k \le q(N-1)$. 
Hence, the implementation of the filling factor implies that only modes with a certain range of momenta exist. In particular, 
this implies that there are no position eigenstates, as they would require all momenta. 

We have derived a one-dimensional limit of the quantum Hall system where the electrons move on a circle and the hamiltonian consists of an electron-electron 
interaction only---there is no kinetic term. This defines our one-dimensional theory, the quantum Hall circle. 
It is interesting to note that the model arrived at here as a limit of the quantum Hall system is, in fact,  identical  to the toy model proposed by Dyakonov \cite{Dyakonov}. 

\section{Quantum Hall states}\label{qh}

In this section we consider wave functions for the QH circle. We relate each bulk QH wave function to a unique circle wave function. Furthermore, we establish that the Laughlin state is, 
for a short range interaction, the ground state on the QH circle  and that it develops adiabatically into the bulk Laughlin state as $L$ decreases. 

We start by taking the QH circle limit of bulk wave functions. For definiteness, we consider the Laughlin states, but  the discussion is general and extends to, for example,  non-abelian states.
The Laughlin wave function at filling fraction $\nu=1/q$ on the cylinder reads  \cite{thouless}
\begin{equation}\label{laughlin}
\Psi_{1/q}=\prod_{i<j}(e^{2\pi i z_i/L}-e^{2\pi i z_j/L})^q \ 
e^{-\frac 1 2 \sum_n y^2_n}
\end{equation}
where $z_i=x_i+i y_i$. $\Psi_{1/q}$ is the exact ground state for a certain short range interaction and there is a gap to all excitations \cite{haldanebook,trugman,pokrovsky}. Since this result follows from the short distance properties of the wave function it holds also on a cylinder for any  
(finite) $L$ \cite{Haldane94}; an explicit proof of this can be found 
in Ref. \cite{bk3}.

Expanding  $\Psi_{1/q}$
in powers of $e^{2\pi i z/L}$ and using that the single particle
states \pref{psicyl} can be written as
$\psi_k=\pi^{-1/4}L^{-1/2}(e^{2\pi i z/L})^k
e^{-y^2/2}e^{-2\pi^2k^2/L^2}$, one finds
\begin{eqnarray}\label{laughlinexpand}
 \Psi_{1/q}&=&\sum_{\{k_j\}}
c_{k_1\dots k_N}\prod_j(e^{2\pi i z_j/L})^{k_j}e^{-\frac 1 2 \sum_n
y^2_n} \\&=&\!\pi^{N/4}\!L^{N/2}\!\sum_{\{k_j\}}
\!c_{k_1\dots k_N}e^{2 \pi^2
\sum_n k^2_n/L^2}\!\prod_j\psi_{k_j}(z_j) \ , \nonumber
\end{eqnarray}
where $c_{k_1\dots k_N}$ are anti-symmetric, integer-valued
coefficients independent of $L$ and the total momentum $\sum_ik_i$ is the same for each set $\{k_j\}$. The weight of
a particular electron configuration, {\it ie} of a Slater determinant, is multiplied by the factor
$e^{2 \pi^2 \sum_m k^2_m/L^2}$ \cite{Haldane94}. In the Tao-Thouless limit, $L\rightarrow 0$, the configuration with maximal $\sum_m k^2_m$ dominates; the electrons are then situated as far apart as
possible along the cylinder, for the Laughlin state at every $q$:th site. The QH circle limit is obtained by letting $L \rightarrow \infty$ in  \pref{laughlin}, this gives 
\be{gs}
\Psi_{L\rightarrow \infty} = \prod_{i<j} (e^{i\varphi_i }-e^{i \varphi_j})^q  \, e^{-\frac 1 2\sum_n y_n^2} \  .
\ee
Dropping the gaussian $e^{-\frac 1 2 \sum_n y_n^2}$, which  has no relevance in the one-dimensional setting, the state on the QH circle is obtained.
The Laughlin state approaches this state continuously---in occupation space, according to \pref{laughlinexpand}, the weights $c_{k_1\dots k_N}e^{2 \pi^2
\sum_n k^2_n/L^2}$ smoothly approach $c_{k_1\dots k_N}$  and $\psi_k \rightarrow e^{ik\varphi}e^{-y^2/2}$. This establishes that the Laughlin state is continuously connected to a corresponding state on the QH circle. The expansion in \pref{laughlinexpand}, with $L$-independent coefficients $c_{k_1\dots k_N}$, holds for any state in the lowest Landau level 
since the latter is a polynomial in $e^{2\pi iz_j/L}$. Thus, \pref{laughlinexpand} provides an explicit  one-to-one correspondence between an arbitrary bulk state and a corresponding circle state. The same Slater determinants, {\it ie}, the same coefficients $c_{k_1\dots k_N}$,  are present in the bulk and on the circle.

To investigate the ground state on the QH circle we consider the effective interaction $\tilde V(\varphi)$ corresponding to a two-dimensional short range interaction 
$V({\bf r})$ \cite{haldanebook,trugman,pokrovsky}. Following the approach of Trugman and Kivelson \cite{trugman}, we have
\begin{equation}
V(\mathbf r)=\sum_{s=0}^{\infty}V_s(\mathbf r)=\sum_{s=0}^{\infty}
c_s b^{2s}\nabla^{2s}\delta_p(\mathbf{r})\ \ ,\label{cylint}
\end{equation}
where $b$ is the range of the interaction, $c_s$ are positive constants, and
$\delta_p(x,y)=\sum_n\delta(x+nL,y)$ is the periodic
delta-function. The  leading term, $\expect{V_0}$, is identically zero for any
fermionic state. If $b\rightarrow 0$, only the leading non-vanishing term in \pref{cylint} contributes to the energy
$E=\expect{V(\mathbf r)}$.

A straightforward calculation, using \pref{veff} for the lowest Landau level, gives the effective one-dimensional interaction
\be{vtwiddle1}
\tilde V(\varphi)&=&\sum_{s=0}^{\infty}\tilde V_s(\varphi) 
=\frac 1 {\sqrt {2 \pi }}\sum_{s=0}^\infty c_s b^{2s} \\
&\times &\sum_{k=0}^s (\frac {2\pi} L)^{2(s-k)+1}(-1)^k(2k-1)!! {s \choose k}  \partial_\varphi^{2(s-k)}\delta (\varphi)  \nonumber \ .
\ee
Keeping the leading term in $b$ for each order of derivatives, \pref{vtwiddle1} becomes
\be{seffint}
\tilde V(\varphi)=\frac {\sqrt{2\pi}} L \sum_{s=0}^{\infty} c_s (\frac {2\pi  b} L)^{2s}  \partial_\varphi^{2s}\delta (\varphi) \ ,
\ee
which, using  \pref{v1d},  gives the matrix elements
\be{v1dkm}
V_{km}=\frac 1 {\sqrt{2\pi} L}\sum_{s=0}^\infty c_s (\frac {2\pi b} L)^{2s} (-1)^{s+1}(k^{2s}-m^{2s}) \ .
\ee

For the interaction $\tilde V(\varphi)$ in \pref{seffint} the argument given in Ref.  \cite{trugman} applies, see \ref{appA}. 
However, the interaction \pref{seffint} is in fact not the 
$L \rightarrow \infty $ limit of the two-dimensional short-range interaction \pref{cylint}. The latter 
leads, for general $L$ in the lowest Landau level, to the following cylinder matrix elements for $V_s({\bf r})$:
\be{vkms}
V_{km}^s&=&\frac {c_sb^{2s}} {L}  \sqrt{\frac 2 \pi} e^{-2\pi^2 (k^2+m^2)/L^2}  (\frac {2\pi} L)^2 (k^2-m^2) v_{km}^s\  , \nonumber 
\ee
where
\be{vkmgen}
v_{km}^1&=& 1\\
v_{km}^2&=& -4 \nonumber \\
v_{km}^3&=& 27- 6(\frac {2\pi} L)^2 (k^2+m^2)
+(\frac {2\pi} L)^4(k^2-m^2)^2  \nonumber \\
v_{km}^4&=& -240+96 (\frac {2\pi} L)^2(k^2 + m^2) 
- 16  (\frac {2\pi} L)^4 (k^2-m^2) ^2  \nonumber \ .
\ee
This does not reduce to \pref{v1dkm} for large $L$. For $s=1$ the same term as in \pref{v1dkm} is obtained but with a modified coefficient, whereas for $s>1$ different terms are obtained. The reason for this discrepancy is that the limit \pref{psilimit} is not consistent for the short-range interaction \pref{cylint}; taking the limit as in \pref{v1d}, using \pref{psilimit}, misses terms obtained when partially integrating the $y$-derivatives onto the gaussian factors.   

Consider an interaction on the QH circle consisting of the $s=1$ term  in (\ref{seffint},\ref{v1dkm})  only. This is a non-negative operator and one finds that the zero energy eigenstates are the states that contain at least three Jastrow factors: $\Psi = fJ^3$, where $J=\prod_{i<j}(e^{i\varphi_i}-e^{i\varphi_j})$ and $f$ is a symmetric polynomial, see \ref{appA}.  All other states have a finite positive energy. At filling factor $\nu=1/3$, the restriction on the momenta, $0\le k \le 3(N-1)$, implies that there is only one zero energy state: The Laughlin state $\Psi = J^3$ is the unique ground state.  This is true also on the cylinder for arbitrary finite $L$ with the hamiltonian in \pref{vkms} (with $J$ being the two-dimensional Jastrow factor). Since  \pref{v1dkm}, with $s=1$, is the $L\rightarrow \infty$ limit of \pref{vkms} (up to a trivial rescaling) it follows that for $\nu=1/3$ the ground state on the QH circle develops adiabatically into the bulk Laughlin state as $L$ grows.

The Laughlin state is the unique ground state to a short range hamiltonian on the QH circle also at filling factors $\nu =1/q< 1/3$ and is adiabatically connected to the bulk Laughlin state. However,  to establish this requires,  for technical reasons, a bit more care; we refer to \ref{appA} for a discussion of this case. 

A comment on the gapless edge excitations present on the cylinder for finite $L$ is in order. One might worry that these survive when $L\rightarrow \infty$ and lead to gapless excitations on the QH circle, thus destroying the argument above.  This is not the case provided the momenta $k$ are restricted so that the edge excitations are excluded; for the Laughlin state at $\nu =1/q$ this is achieved if $0\le k \le q(N-1)$. 
However, if the restriction on the momenta is loosened then gapless excitations will appear in the one-dimensional theory \cite{Haldane94}; just as edge excitations appear on the cylinder.

\section{Crystal}\label{crystal}

We saw above that since the range of momenta is restricted by the filling factor it is not possible to construct position eigenstates and, moreover, it is this fact that 
makes the model non-trivial. We here follow Dyakonov \cite{Dyakonov} and introduce a Wannier-like basis of states, $\chi_s$,  localized around fixed positions on the circle:
\be{wannier}
\chi_s(\varphi) &=&\frac 1 {\sqrt{M}} \sum_{k=0}^{M-1}\psi_k(\varphi)e^{-ik(2\pi s/M+\alpha)}\nonumber \\
&=&\frac 1 {\sqrt{2\pi M}} \sum_{k=0}^{M-1}e^{ik(\varphi-2\pi s/M-\alpha)}\\
\psi_k (\varphi)&=&\frac 1 {\sqrt{M}} \sum_{s=0}^{M-1}\chi_s(\varphi)e^{ik(2\pi s/M+\alpha)}\nonumber \\
&=&\frac 1 {\sqrt{2 \pi}} e^{ik\varphi} \ , \
\ee
where $k,s=0,1,\dots M-1$ and $\alpha$ is  a constant. The wave-functions $\chi_s(\varphi)$ are peaked at $\varphi=2\pi s /M+\alpha$ with a 
width $\Delta \varphi \sim 1/M$ since $\Delta k=M$.

If the average distance, $2\pi  /N$, between two nearby electrons on the circle is much larger than the width of the localized single-particle states $\chi_s$, {\it ie} if 
$2\pi/ N\gg 1/M$ or, equivalently, $\nu=N/M \ll 1$, then the ground state is presumably, as suggested in Ref \cite{Dyakonov}, a crystal-like state where the electrons are as far 
separated as possible, {\it ie} a Slater determinant of Wannier states $\chi_s (\varphi)$.\footnote{Order may exist over a long range, even though true long range order is destroyed by quantum fluctuations.}
On the other hand, if the inter-particle distances are comparable to the width of the most localized states then the interaction may lead to more complicated ground states, such as
the Laughlin state. The crystal identified here on the circle for small $\nu$ breaks spontaneously
the translation symmetry present in the hamiltonian \pref{hamcont1d} by having a fixed parameter $\alpha$ in \pref{wannier} and hence it is a state with gapless excitations, phonons. 
We suggest that this state is adiabatically connected 
to the Wigner crystal expected in the quantum Hall regime for low enough filling factor \cite{w0}. 

This gapless Wigner crystal on the QH circle must not be confused with the gapped 
Tao-Thouless state which is the ground state as $L\rightarrow 0$ and is adiabatically connected to an abelian bulk QH state.  Both have a crystalline structure: The Wigner 
crystal is a single Slater determinant of Wannier states $\chi_s$ localized in the direction around the cylinder, whereas the Tao-Thouless state is a single Slater determinant of 
states $\psi_k$ localized along the cylinder. In each case the 
single-particle states are as far apart as possible in order to minimize the electrostatic repulsion. The two states may seem similar but are in fact unrelated: 
The former is a gapless crystal whereas the later is a gapped quantum Hall state.

\section{Discussion}\label{disc}

We have introduced the quantum Hall circle as the one-dimensional limit of the quantum Hall problem on a cylinder when the circumference goes to infinity and established a one to one correspondence between the states on the circle and the cylinder.  For 
filling fraction $\nu=1/q$ and a short-range interaction, the limit of the Laughlin wave function is found to be the ground state and it has a gap to all excitations. This ground state is 
adiabatically connected to the bulk Laughlin state and, when the circumference goes to zero, to the Tao-Thouless state. We suggest that this analysis generalizes to more general QH states, including states with non-abelian excitations, which are ground states of some short range interaction. For very small filling factors, the QH circle is argued to be a gapless state related of the Wigner crystal expected in the bulk.

It is not obvious that the quantum Hall circle is any easier to analyze than the full QH problem; the hamiltonian is still given by Eq. \ref{hammom}  but with a special choice of $V_{km}$ that 
do not look particularly simple. 
The situation differs from the opposite, Tao-Thouless, limit, where only the electrostatic terms $V_{k0}$ survive and, 
as a consequence, the hamiltonian can be diagonalized. For the quantum Hall circle, the full complexity seems to remain. A potentially useful feature is, however, that 
from a wave function on the circle the corresponding bulk wave function is obtained simply by the replacement $\frac 1 {\sqrt{2\pi}} e^{ik\varphi} =\psi_k(\varphi) \rightarrow (e^{2\pi i z/L})^k e^{-y^2/2}=  e^{2 \pi^2 k^2/L^2}\psi_k({\bf r})$, where $\psi_k({\bf r})$ are the full Landau level wave functions \pref{psicyl};  this for example uniquely gives the full Laughlin, Moore-Read or Read-Rezayi wave function from 
its QH circle limit.\footnote{Clearly this replacement can also be used in other geometries (and gauges); in the disk (symmetric gauge) it would be $\psi_k(\varphi) \rightarrow z^ke^{-|z|^2/4}= \sqrt{2^{k+1}\pi k!}\psi_k({\bf r})$.}

The interest in the QH circle is at this stage perhaps mostly conceptual in that it provides a one-dimensional limit where the QH physics remains. This may lead to useful connections to other one-dimensional systems and the powerful methods developed for such systems may perhaps be applied here as well. In this context, we would like to point to the similarity with the Calogero-Sutherland models \cite{fc,bs} and, in particular, to the Haldane-Shastry spin chain where the ground state is of Laughlin type \cite{hs1,hs2}.

\ack
We gratefully acknowledge  discussions with Thors Hans Hansson and Susanne Viefers.
AK was supported by the Swedish Research Council and by NordForsk.

\appendix

\section{Ground states on circle}\label{appA}

Here we show that the Laughlin state is the ground state on the QH circle for a short-range interaction and establish the adiabatic continuity from the circle to the cylinder. The argument follows closely Ref. \cite{trugman}.

We have seen that in order to have a fixed filling fraction, 
only single-particle states $\psi_k(\varphi)=\frac 1 {\sqrt{2 \pi}} e^{i k \varphi}$ with momenta $0\le k \le q(N-1)$ are allowed. Moreover, since the hamiltonian \pref{hammom} is translationally 
invariant the energy eigenstates can be chosen as momentum eigenstates. These are states with given $K=\sum _i ^N k_i$ and are 
homogeneous (antisymmetric) polynomials $\Psi$ in $e^{i\varphi}$.  In the bulk the wave functions are holomorphic functions, here this
corresponds to there being no factors $e^{-i\varphi}$; this follows from the momentum restriction on the one-particle states $\psi_k=\frac 1 {\sqrt{2 \pi}}e^{i k \varphi}$. 

The expectation value of  one of the terms in \pref{seffint}, $\partial_\varphi^{2s} \delta (\varphi)$, is an integral over terms $\partial_{\varphi_i} ^{2s} \Psi^* \Psi|_{\varphi_i=\varphi_j }$, $i\neq j$ (with positive coefficients). For $s=1$, this expectation value is non-negative for any state (since $\Psi|_{\varphi_i=\varphi_j }=0$), hence the $s=1$ operator in (\ref{seffint},\ref{v1dkm}) is a non-negative operator and a state is a zero energy eigenstate if and only if its expectation value vanishes; this is the case when $\partial_{\varphi_i} \Psi|_{\varphi_i=\varphi_j }=0$. This  holds if and only if $\Psi \sim (\varphi_i-\varphi_j)^2$ as $\varphi_i-\varphi_j \rightarrow 0$ which is equivalent to  $\Psi =J^2 f$, where  $J=\prod_{i<j} (e^{i\varphi_i }-e^{i \varphi_j})$ is the Jastrow factor and $f$ is a 
homogeneous polynomial in $e^{i\varphi}$. Consider now the expectation value of the $s=2$ operator. If the state is a ground state of the $s=1$ operator, then $\partial_{\varphi_i} ^{4} \Psi^* \Psi|_{\varphi_i=\varphi_j }=|\partial_{\varphi_i} ^{2} \Psi|^2_{\varphi_i=\varphi_j }\ge 0$ and the expectation value is non-negative; it vanishes if and only if $\partial_{\varphi_i} ^{2} \Psi|_{\varphi_i=\varphi_j }=0$ which is equivalent to $\Psi =J^3 f$. Hence,  the $s=2$ operator is non-negative in the space of states $\Psi =J^2 f$ that are the ground states of the $s=1$ operator. However, when diagonalized in the full hilbert space, the $s=2$ operator has negative eigenvalues; as a consequence the states $\Psi =J^3 f$ are not the ground states of a hamiltonian consisting of the $s=1$ and $s=2$ terms in the full hilbert space. This logic continuous to higher $s$ operators in an obvious manner.\footnote{Since the particles are fermions any state must be antisymmetric, hence it contains an odd number of Jastrow factors. Thus, the zero energy states at $s=1$ are actually $\Psi =J^3 f$ just as at $s=2$; it is only at the next level, $s=3$, a restriction on the zero energy states is obtained, in which case they become $\Psi =J^5 f$.} 

What we have presented is just a translation of the argument given by Trugman and Kivelson \cite{trugman} for the two-dimensional QH problem, which applies to the hamiltonian in \pref{vkmgen} for a cylinder of arbitrary radius,  and the conclusions are the same. However, for this case a stronger result in fact seems to hold: The individual operators $\nabla^{2s}\delta_p({\bf r})$ in \pref{cylint} (or $V^s_{km}$ in \pref{vkmgen}) are non-negative in the full hilbert space. We have verified this for the $s\le 4$ on small systems  and it is consistent with the pseudopotential formulation of the short-range interaction \cite{haldanebook}. If this is the case, then the individual terms in \pref{vkmgen} can be used, with finite coefficients, to single out the Laughlin state as a unique ground state, on the cylinder, also at $\nu=1/q<1/3$.

We now show that the Laughlin state is the unique ground state to a short range hamiltonian on the QH circle at filling factors $\nu =1/q$. For $\nu=1/3$, this is true for the hamiltonian $\partial^2_\varphi \delta (\varphi)$ (the $s=1$ operator in \pref{seffint}) provided the momenta are restricted to $0\le k \le 3(N-1)$, so that $\Psi=J^3$ is the only zero energy state.
For $\nu=1/q<1/3$, one needs $s>1$ operators $\partial^{2s}_\varphi \delta (\varphi)$  in the hamiltonian to single out the Laughlin state as the unique zero energy state.
However,  the $s>1$ terms in \pref{v1dkm} are not positive definite and hence the Laughlin state will not be the exact ground state.  A solution to this problem is to restrict the hilbert space to  the zero energy states of $\partial^2_\varphi \delta (\varphi)$, {\it ie}, to the states $\Psi=f J^3$. In this restricted hilbert space the operator $\partial^6_\varphi \delta (\varphi)$  is non-negative with zero energy states $\Psi=fJ^5$.  Restricting the momenta to correspond to $\nu=1/5$, the Laughlin state $\Psi=J^5$ is the unique ground state. The procedure can then be iterated: Restricting to the states $\Psi=fJ^5$, the non-negative operator $\partial^{10}_\varphi \delta (\varphi)$ has the zero energy states $\Psi=fJ^7$, {\it etc}. It should be noted that, when obtaining the Laughlin state $\Psi=J^q$ as the unique ground state at $\nu=1/q$, the restricted hilbert space contains all states with $q-2$ Jastrow factors. This includes the fractionally charged quasiparticles at $1/q$ as well as, we believe, all other low energy excitations---thus we argue that the restriction of the hilbert space does not affect the low energy physics.

We now establish the adiabatic connection between the Laughlin ground state on the QH circle and on the cylinder  by constructing a hamiltonian that interpolates between the two cases while the Laughlin state remains the unique ground state. For $\nu=1/3$ this is trivial since the $L \rightarrow \infty$ limit of \pref{cylint} is $\partial^2_\varphi \delta (\varphi)$.  For  $\nu=1/5$ we proceed as follows; the argument generalizes directly to smaller $\nu=1/q$. As hamiltonian on the circle we take
\be{v1}
V_{1,km}=\lim_{a\rightarrow \infty} a(k^2-m^2)+(k^6-m^6) \ .
\ee
The first term projects onto the states $\Psi=fJ^3$ and in this subspace the second term is non-negative with the zero energy states $\Psi=fJ^5$.\footnote{An alternative to \pref{v1} is $V_{1,km}={\rm lim}_{L\rightarrow \infty}L^7V_{km}$, where $V_{km}$ are given by \pref{v1dkm} with $b$ finite. A term $m^4-k^4$  can be included in \pref{v1} without changing the results.}
Restricting the momenta, the Laughlin state on the circle, {\it ie} the state given by  \pref{laughlinexpand} with $L\rightarrow \infty$,  is the unique ground state. We can now modify \pref{v1} so that the Laughlin state for finite $L$ is still an eigenstate with zero energy---this is achieved simply by the replacement $V_{1,km}\rightarrow e^{-2\pi^2 (k^2+m^2)/L^2}V_{1,km}$. This follows by noting that the exponential $L$-dependence in \pref{vkmgen} is 'universal'; it is the same for all terms and only compensates the corresponding exponential $L$-dependence in the wave function in \pref{laughlinexpand}. Denote this new hamiltonian $H_1(L)$. Since the Laughlin state is the ground state with a gap at $L\rightarrow \infty$ and remains an eigenstate for all $L$ it will, by continuity, be the unique ground state of $H_1(L)$ for large enough finite $L$. 
We introduce the interpolating hamiltonian $H_{\lambda}=(1-\lambda)H_1+\lambda H_2$, where  $H_2$ can be taken as consisting of terms in \pref{vkmgen} (assuming they are non-negative operators) or a construction similar to the one in \pref{v1} can be employed; $\lambda$ goes from
0 to 1 when $L$ goes from $\infty$ to some large but finite value.  From the discussion above
follows that the Laughlin state is the ground state of $H_\lambda$ for all $L$, and that there is a gap to excitations.
This establishes that the QH circle is adiabatically connected to the bulk QH problem, without there being a phase 
transition when one goes from one to the other, for a short range interaction at $\nu=1/q$, $q$ odd.\footnote{Note that rather than changing the size of the cylinder one can think of the process as taking place on 
a cylinder of fixed size changing only the matrix elements $V_{km}$ in the hamiltonian \pref{hammom}. This gives the standard setting for discussing adiabatic continuity.}

\end{document}